\documentclass[nofootinbib,
showpacs,prl,twocolumn,aps,floatfix]{revtex4}

\usepackage{graphicx}

\newcommand {\fexp} [1] {\exp \left( #1 \right)}
\newcommand {\fabsq}[1] {\left| #1 \right|^2}

\newcommand {\fcot} [1] {\cot\left( #1 \right)}

\newcommand {\fsin} [1] {\sin\left( #1 \right)}

\newcommand {\fsinq} [1] {\sin^2\left( #1 \right)}

\newcommand {\fcoshq} [1] {\cosh^2\left( #1 \right)}
\newcommand {\fsinhq} [1] {\sinh^2\left( #1 \right)}
\newcommand {\fcoth} [1] {\coth\left( #1 \right)}

\newcommand {\si} {/\mbox{s}}

\newcommand {\mum}{\, \mu \mbox{m}}
\newcommand {\cms}{\, \mbox{cm/s}}
\newcommand {\cO} {\begin{cal} O \end{cal}}
\newcommand{\beq}{\begin{equation}}
\newcommand{\eeq}{\end{equation}}
\newcommand{\beqa}{\begin{eqnarray}}
\newcommand{\eeqa}{\end{eqnarray}}

\begin{document}
\title{Velocity selection 
of ultra-cold atoms with Fabry-Perot laser devices:
improvements and limits}

\author {A. Ruschhaupt}

\author {F. Delgado}

\author {J. G. Muga}
\affiliation{Departamento de Qu\'{\i}mica-F\'{\i}sica,
UPV-EHU,\\
Apartado 644, 48080 Bilbao, Spain}

\begin{abstract}
We discuss a method to select the velocities of ultra-cold atoms
with a modified Fabry-Perot type of device made of two
effective barriers and a well created, respectively, by 
blue and red detuned lasers.   
The laser parameters may be used to 
select the peak and width of the transmitted
velocity window. In particular, lowering the central 
well provides a peak arbitrarily close to zero velocity  
having a minimum but finite width.   
The low-energy atomic scattering off this laser device
is parameterized 
and approximate formulae are 
found to describe and explain its behaviour.  
\end{abstract}
\maketitle

Velocity selection 
is a basic operation 
in quantum optics and atomic physics for a plethora of
applications. 
There are mechanical (slotted
disks) and non-mechanical (optical) 
techniques available, useful for different
experimental circumstances, species, and energies.  
The large wavelengths achieved with laser cooling
have made the traditional methods no longer effective  
because 
of the increasing importance of gravity
and the  
quantum nature of translational motion. 
For example, the standard 
classical-mechanical analysis of mechanical velocity-selection
methods becomes 
invalid for small-time temporal slits, since they 
produce momentum spread in agreement with  
a time-energy uncertainty principle \cite{David}. 
Among the new methods, the velocity selection using
Doppler sensitive stimulated Raman transitions \cite{chu2}, 
and coherent population trapping
into a dark state \cite{Aspect}, provide selectivity in the 
``transverse direction'' parallel to the 
lasers, and rely on specific internal level configurations. 
\begin{figure}
\begin{center}
\includegraphics[width=.8\linewidth]{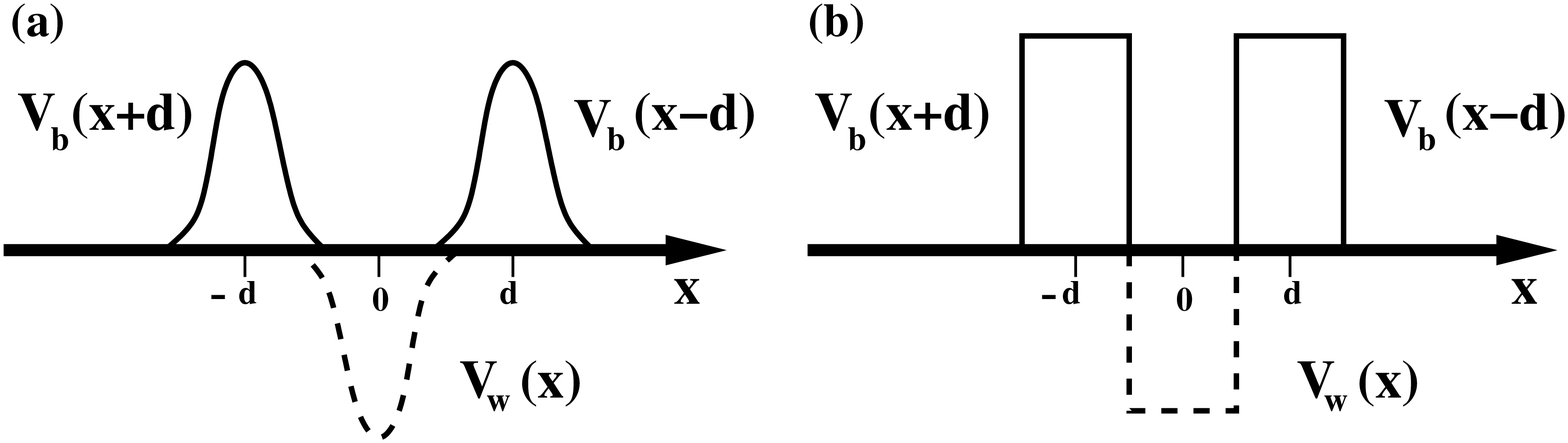}
\end{center}
\caption{\label{fig1} $m=\mbox{mass}({ }^{23}\mbox{Na})$;
(a) Gaussian functions, $d = 6 \mum$, $\sigma = 2 \mum$;
(b) square functions, $d = 5 \mum$}
\end{figure}
%
Fabry-Perot (FP) cavities have been also proposed to provide coherent 
velocity selection or trapping for longitudinal motion, 
using detuned lasers perpendicular to the incident atoms \cite{WGTM93}
or microwave cavities \cite{LMW98,2cav}
to implement the 
partially reflecting mirrors. 
The velocity selection in these 
cavities is produced by the filtering effect of
resonance peaks in the
transmission probability. The potential of FP cavities
as trapping devices 
also stems from characteristic resonance features: 
high densities and large 
life times in the
interaction region.    

\begin{figure}[t]
\begin{center}
\includegraphics[angle=-90,width=1.0\linewidth]{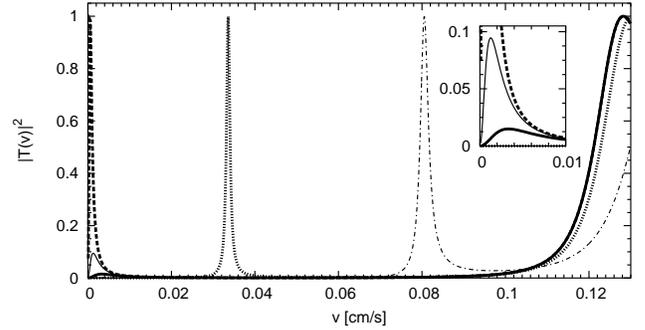}
\end{center}
\caption{\label{fig2}Transmittance versus velocity for 
different well
depths. $\widehat{V}_{b}=300 \,\hbar\si$;
$\widehat{V}_w = 0$ (dashed-dotted line), $\widehat{V}_w = 150\,\hbar\si$
(thick dotted line),
$\widehat{V}_w = 180.2\,\hbar\si$ (thick dashed line), $\widehat{V}_w = 
180.25\,\hbar\si$
(solid line), $\widehat{V}_w = 180.5\,\hbar\si$ (thick solid line);
other parameters in Fig. \ref{fig1}a. The inset is a zoom
of the lower-left corner.}
\end{figure}
The aim of this work is to discuss an improvement 
of these cavities, provide formulae to describe their
behaviour, 
and study the fundamental limitations to lower 
the peak width and velocity of the transmitted wave packet. 
The basic idea is to add a well with controllable depth 
between the two external barriers, see Fig. 1.
Effective barriers and well can
be implemented 
with blue and red detuned lasers, respectively, which do not excite 
the impinging ground state atom and cause only a mechanical effect.
The depth of the well and the barrier height can be varied with the 
intensities of the lasers. 
Making the well deeper, rather than wider, 
displaces the resonance peaks to lower energies
without diminishing the inter-resonance spacing  
so it is the ideal way to achieve a sharp low-energy velocity selection. 
The resonance peak displacement with the well-depth can be 
seen in Fig. 2. The velocity shift is accompanied by a peak width reduction 
until a minimum, non-zero width is attained when the peak reaches zero 
velocity at a critical ``threshold'' depth. Beyond that
depth the peak broadens, moves to higher velocities, and
its maximum decays, as shown in the inset of Fig. 2.
The effects of different depths are summarized in Fig. 3, 
which will be 
explained next in more detail.

We shall use both a realistic model based on three Gaussians, see 
Fig. 1, as well as a simplified version with two square barriers 
and a well. The scattering off the two potential models is very similar  
but the later  
enables us to obtain analytical exact results and approximate but 
physically illuminating expressions. 
%
%
\begin{figure}[t]
\begin{center}
\includegraphics[angle=-90,width=1.\linewidth]{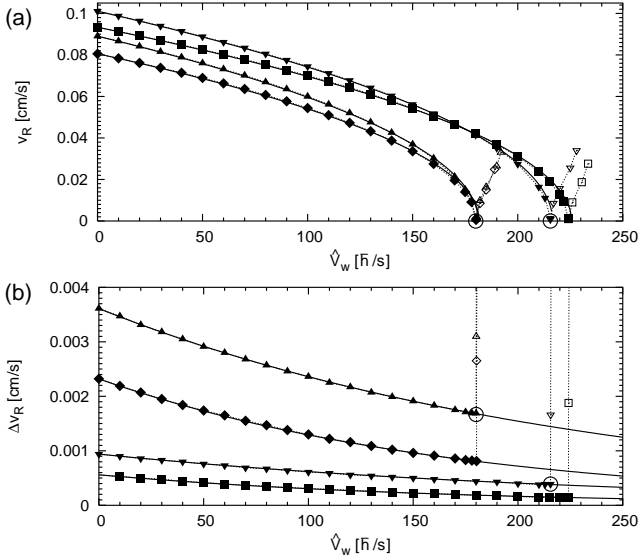}
\end{center}
\caption{\label{fig3}
(a) Resonance velocity $v_R$ versus $\hat{V}_w$  
(symbols connected with dotted line).
Filled symbols indicate the case $\fabsq{T(v_R)} > 0.995$, and empty 
symbols
otherwise. The solid
lines show the approximation of Eq. (\ref{approx1}).
Gaussian functions, see Fig. \ref{fig1}a:
$\widehat{V}_b=300 \hbar\si$, $\alpha=0.65$ (diamonds),
$\widehat{V}_b=500 \hbar\si$, $\alpha=0.70$ (squares);
square functions, see Fig. \ref{fig1}b:
$\widehat{V}_b=300 \hbar\si$, $\alpha=0.79$ (triangles up),
$\widehat{V}_b=500 \hbar\si$, $\alpha=0.85$ (triangles down),
the circles indicate $\widehat{V}_{w,thres}$ for the square model 
in the 2-pole
approximation.
(b)Velocity width $\Delta v_R$ of the resonance versus
$\widehat{V}_w$; meaning of symbols as in (a).
The solid lines show the approximation of Eq. (\ref{delv}),
the circles indicate $(\widehat{V}_{w,thres},
{\Delta v}_{R,thres})$ with the 2-pole approximation.}
\end{figure}
%
Let us consider, for a single ultra-cold atom,
the Hamiltonian
\begin{eqnarray}
H = -\frac{\hbar^2}{2m}\,\frac{\partial^2}{\partial x^2} 
+  V_b (x+d) - V_w (x) + V_b (x-d),
\label{ham}
\end{eqnarray}
where $V_{b,w} (x)= \widehat{V}_{b,w} \;\Pi(x)$, and 
$\Pi$ can take the forms 
\begin{eqnarray*}
\Pi_G\!=\!\fexp{-\frac{x^2}{2 \sigma^2}},\;\;
\Pi_s\!=\!\left\{
\begin{array}{cc}
1&{\rm if} -d/2<x<d/2
\\
0&{\rm otherwise}
\end{array}
\right. 
\end{eqnarray*}
for the Gaussian and square models respectively. 
For simplicity we have set all Gaussians with the same width $\sigma$, and 
the square 
segments with the same length $d$.  
We assume  
that the atom impinges from the left and only initial 
positive velocities are considered. 
We are interested in the 
transmission amplitude $T$ 
and the ``transmittance'' $|T|^2$ 
of the scattering solutions of 
$
H \phi_v (x) = E_v \phi_v (x),  
$
where $E_v = \frac{m v^2}{2}=\hbar^2 k^2/(2m)$.
Both velocity, $v$, and wavenumber, $k$, will be used,
the later being more appropriate for complex plane analysis and the former 
for presenting the physical results.     
For the square model,
%
%
\begin{eqnarray}
T(k)&=& -4 e^{-2idk} k k_b^2 k_w 
\label{trans}
\\
&\times&\Big\{ e^{id(k- k_w)}\left[
i k_b  (k+ k_w) C  
+ (k k_w-k_b^2) S 
\right]^2
\nonumber\\
&\!-\!&\!\!e^{id(k+ k_w)}\!\!
\left[
-i k_b (k- k_w) C 
\!+\! (k k_w+k_b^2) S 
\right]^2\!\!\Big\}^{\!-1}
\nonumber
\end{eqnarray}
%
where  
$C=\cosh(d k_b)$, $S=\sinh(d k_b)$, 
$k_w=(k^2+K_w^2)^{1/2}$,
$k_b=(K_b^2-k^2)^{1/2}$, and 
$K_b=(2m\widehat{V}_b)^{1/2}/\hbar$, 
$K_w=(2m\widehat{V}_w)^{1/2}/\hbar$. 

\begin{figure}[t]
\begin{center}
\includegraphics[width=3.5cm,height=7cm,angle=-90]{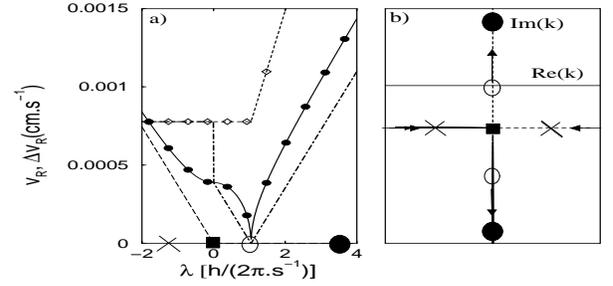}
\end{center}
\caption{\label{fig4} Connection between peaks and poles in the square model,
$\widehat{V}_b=400\hbar\si$, other parameters in Fig. \ref{fig1}b;
for the plotted parameter range, the 2-pole approximation and
the exact result are indistinguishable;
(a) resonance velocity $v_R$ (line with filled dots),
resonance width $\Delta v_R$ (line with unfilled diamonds)
versus the distance $\pm \lambda$ of the pole to the collision point
($\lambda < 0$ before the collision, $\lambda > 0$ after the
collision); a 1-pole approximation with only $k_1$ is also plotted
($v_R$ (dashed line), $\Delta v_R$ (dashed-dotted line)).
(b) Motion of the two poles:
$\widehat{V}_w=199.884\hbar\si$ (crosses),
$\widehat{V}_{w,coll}=199.898\hbar\si$ (coinciding squares),
$\widehat{V}_{w,thres}=199.901\hbar\si$ (small circles),
$\widehat{V}_w=199.92\hbar\si$ (big circles).}
\end{figure}

In both models, at some critical,
``threshold'' well-depths $\widehat{V}_{w,thres}$ 
new bound states are formed,  
and for well-depths close to these thresholds,
the description of the transmission 
peak is not as simple 
as for isolated Breit-Wigner (BW) resonances (see Fig. \ref{fig4}). At a depth
$\widehat{V}_{w,coll}$ 
slightly before threshold, 
a resonance-antiresonance pole collision occurs, 
and with further deepening, two ``virtual'' states in the complex momentum
plane appear. 
Centering our attention on the first, lowest-energy resonance at zero  
well-depth, one can distinguish 
when increasing the well-depth:
first an ordinary resonance regime with a  
BW transmittance peak; second, an intermediate pre-bound state regime  
near threshold, in which the second pole cannot be ignored;
and finally a bound-state regime. In Figure \ref{fig2} transmittance curves 
corresponding to the different regimes are depicted. 
Even though the calculations can be made exactly, it is 
useful for applications and physically illuminating to 
describe these three stages in terms of approximate expressions   
and dominant dependences relating:
poles of $T$ in the complex momentum plane;  
well-depth or other potential parameters;
and visible features such as position and width of the
resonance.

The first stage, dominated by a BW resonance pole $k_1$      
in the fourth quadrant of the momentum complex plane, 
is the most important one for 
velocity selection since it allows to diminish the resonance velocity 
and width by deepening the well, see Figs. \ref{fig2} and \ref{fig3}a.    
$k_1$ is accompanied by an antiresonance at 
$k_2=-k_1^*$.
Since the barrier is symmetrical with respect to parity,
the (Gamow) resonance states 
have well defined parity. 
Thus they appear alternatively in one of the two eigenvalues 
of the $2\times2$ $S$ matrix, for symmetrical, 
$S_0$, or antisymmetrical scattering, $S_1$.
The transmission amplitude
is given by 
$
T=(S_{0}+S_{1})/2. 
$
We shall follow the 
motion of the first symmetrical resonance, 
the one that will become the ground state 
for deep enough wells,  
and assume that the first antisymmetrical pole 
of $S_1$ is far from the origin so that a 
two pole approximation suffices. Retaining only two poles 
in the canonical pole expansion for cut-off potentials, 
\beq
S_0=-e^{-2ikr}\frac{(k-k_1^*)(k-k_2^*)}{(k-k_1)(k-k_2)},\;\;\;
S_1=e^{-2ikr}.
\label{ss1}
\eeq
Here $r$ is $3d/2$ for the square model. For the Gaussian model, 
we could  
truncate the potential at a large $r$ value and apply  
Eq. (\ref{ss1}).
In any case 
the phase factor does not play any role to calculate the filtering 
function $|T|^2$. 

Expressions for the two important poles can be obtained with   
the square model 
under some approximations, as we shall see later on.  
In the ``BW'' regime 
the antiresonance $k_2$ may normally be ignored if the resonance is 
sharp (i.e. $k_1$ is close to the real axis) and far from the origin.    
A decrease in the well-depth displaces $k_1$ to the left
and upwards, 
so that the transmittance curve decreases both its peak velocity and 
width. By inspection of the $S$ matrix, 
it is clear that in this regime 
the transmittance reaches the unitary limit $|T|^2=1$
close to $\Re {\rm{e}} (k_1)$.  
We define $E_R=mv_R^2/2$ and $v_R$ as the energy and velocity 
of the transmittance maximum. 
In the BW regime $E_R\approx \hbar^2\Re {\rm{e}}(k_1)^2/2m$.
Simple parameterizations of this regime are provided  
by perturbation or semiclassical formulae. 
Let $E_{R0}$ and $v_{R0}$ be the real energy and velocity 
of the resonance peak ``without
well'' ($\widehat{V}_w=0$).
Then the energy of the resonance with
non-zero well can be approximated within a perturbation theory for
resonance functions \cite{kukulin.book}. 
Up to first order in $\widehat{V}_{w}$,
\beq
E_R=E_{R0} - \alpha \widehat{V}_w,\;\;\;
v_R = \sqrt{v_{R0}^2 - 2\alpha \widehat{V}_w/m}.
\label{approx1}  
\eeq
%
A semiclassical treatment \cite{Bohm} for opaque 
barriers gives $\alpha=1$, 
but keeping $\alpha$ as a fitting parameter the dependence 
of Eq. (\ref{approx1}) is valid even beyond very opaque
barriers or very small depths,
as can be seen in Fig. \ref{fig3}a. 
We define a velocity width $\Delta v_R$ as the width of the transmittance 
peak at 
half height. In 
Fig. \ref{fig3}b the BW regime corresponds to the slow decrease with 
well-depth up to 
the abrupt, almost vertical increase associated with a bound state.     
A semiclassical estimate for the energy-width of the resonance is 
given by 
a well-frequency factor times 
the WKB probability to escape through a barrier from the well
(see e.g. \cite{Bohm}). Retaining dominant dependences in the 
opaque and shallow well limit,  
\beq
\Delta v_R= \Delta v_{R0} \exp(-\beta \widehat{V}_w),
\label{delv}
\eeq
which, again, by keeping $\beta$ as an effective fitting parameter, 
describes the correct behaviour in the whole BW regime, until 
well-depths very near the intermediate threshold region, see Fig. \ref{fig3}b.  
%
%

Near the threshold depth 
the velocity and width of the peak are
affected more and more 
by the nearby 
antiresonance, $k_2=-k_1^*$.
This intermediate regime is extremely narrow, with respect to 
variations of $\widehat{V}_w$, compared to the BW
and bound state ones (see Fig. \ref{fig4}). Nevertheless,  
its analysis is worthwhile since it establishes the ultimate 
physical lower limit of the peak velocity and width using a 
FP filtering device. 
At a critical ``collision'' depth $\widehat{V}_{w,coll}$ 
both poles meet at $-i\kappa_{coll}$, $\kappa_{coll}>0$,  
on the negative imaginary axis (see e.g. the ``square'' in Fig. \ref{fig4};
in one dimensional scattering,  
as for s-wave scattering, the collision is not at the origin
because bound states are not degenerate).   
Note that, in spite of their zero real part, 
the velocity peak is not at $v=0$.  
As the well becomes more profound the two poles  move in opposite 
directions, now along the imaginary axis as ``virtual''  poles
until the upper one arrives at the origin at the threshold
depth $\widehat{V}_{w,thres}$, with the lower pole at $-i\kappa_{thres}$
(see e.g. ``circles'' in Fig. \ref{fig4}).
The motion of the two poles just before the collision and even beyond threshold 
is well described by expanding the denominator 
of $S_0$ in powers of $(\widehat{V}_w-\widehat{V}_{w,coll})$
and retaining the first term. This gives  
%
$
k_{1,2}=-i\kappa_{coll}
\pm i \gamma (\widehat{V}_{w}-\widehat{V}_{w,coll})^{1/2},
$
%
with $\gamma$ real. A consequence is that, at threshold,  
$k_1=0$, $k_2=-i\kappa_{thres}\approx -2i\kappa_{coll}$.
The threshold 
is a singular, abnormal  
point
in which the 
transmission peak reaches the origin, 
$T(0)=1$ ($T(0)=0$ for any other well-depth).
Moreover, from the 2-pole approximation of $S$,  
%
$
|T|^2={\kappa_{thres}^2}/({k^2+\kappa_{thres}^2}). 
$
%
Thus, the width at half height, considering
only positive momenta, reaches its minimum value.   
In wavenumber units it is just $\kappa_{thres}$,
and the maximum, $|T|^2=1$,
occurs at  $k=\kappa_{thres}$.
Approximate expressions for $\kappa_{thres}$ may be obtained from 
Eq. (\ref{trans}).  
We have to find zeros of the denominator of $T$. 
Let $\chi \equiv \frac{k}{K_b}$. 
We assume $\widehat{V}_b,\widehat{V}_w > 0$ and $\chi \ll 1$.
Neglecting $\cO(\chi^3)$ we arrive at the quadratic equation
%
$\alpha_2 \chi^2 +2i \alpha_1 \chi - \alpha_0 = 0$
%
with $\alpha_0 = \fcot{\frac{d}{2} K_w} - K_w\fcoth{d K_b}/K_b$,
$\alpha_1 = K_w/\left(2 K_b \fsinhq{d K_b}\right)$, and
\begin{eqnarray*}
\alpha_2 &=& \frac{K_b^2\left(d K_w + \fsin{d K_w}\right)}
{4 K_w^2 \fsinq{d K_w/2}}\\
& & + \frac{K_w\left(\fcoth{d K_b} \left(\fcoshq{d K_b} - 3\right) + d K_b\right)}
{2 K_b \fsinhq{d K_b}},\\
\end{eqnarray*}
%
The two solutions are given by
$\chi_{1/2} =
-i \alpha_1/\alpha_2 \pm \sqrt{\alpha_0\alpha_2-\alpha_1^2}/\alpha_2$.
At $\widehat{V}_{w,thres}\equiv\hbar^2K_{w,thres}^2/2m$,
a zero of the denominator of $T$ is at $k=0$, so  
$\alpha_0 = 0$. Note that if $K_b\gg 1$, 
$K_{w,thres}\approx\pi/d$.
The other pole is at $-i\kappa_{thres}=-2iK_b\alpha_1/\alpha_2$
and determines the minimal velocity width 
${\Delta v}_{R,thres} = \hbar\kappa_{thres}/m$
(see Fig. \ref{fig3}).

Finally, with further well deepening, the upper pole 
crosses the real axis and becomes a bound state. 
As predicted by Eq. (\ref{ss1}),  
the transmittance peak broadens dramatically
and moves to higher positive velocities; also  
the peak maximum becomes smaller than one, so this 
regime is no longer useful for velocity filtering, 
see Figs. \ref{fig2} and \ref{fig3}.

As an application example
we shall compute  
the transmitted velocity distributions resulting from the FP 
filtering of 
an an atomic wavepacket prepared as a  
Bose-Einstein condensate \cite{anglin.2002,bongs.2004}
in a trap.
The trap is moved with a certain velocity with respect to 
the laboratory frame and turned off suddenly at $t=0$.  
The condensate expands until the nonlinear interaction
between the atoms can be neglected and encounters the FP cavity.  
(Alternatively the triple potential can be moved
with the trap at rest.)
First we calculate numerically the ground state $\bar{\psi}_0$
(normalized to $1$) in the reference frame where the trap is at rest.
A harmonic trap is assumed with frequency $\omega_x$ in $x$ direction
and $\omega_{yz}$ in $y$ and $z$ directions.
Using a one-dimensional approximation
of the Gross-Pitaevskii equation the Hamiltonian is 
$ H(\bar{\psi})\!=\!\frac{-\hbar^2}{2m}
\frac{\partial^2}{\partial x^2}\!+\!
\frac{m\omega_x^2}{2}x^2
\!+\! 2\hbar N a
\omega_{yz} {\fabsq{\bar{\psi} (x)}},$
%
where $N$ is the number of atoms in the condensate
and $a$ the scattering length. We take  $a=2.93\times 10^{-9} \, m$. 
Then we change  
to the lab frame where the trap moves with velocity
$v_0$. The ground state $\psi_0$ in this reference frame is 
$\psi_0 (x) = e^{i x \frac{mv_0}{\hbar}} \bar{\psi}_0 (x)$.
At $t=0$ the trap is turned off being at position $x_{TRAP}$
and the velocity selection potentials are
switched on, i.e., the time-evolution is
given by the Gaussian version of the Hamiltonian of 
Eq. (\ref{ham}) plus a term $V'$ 
representing the decaying 
non-linear effect due to free expansion in $y$ and $z$ directions,
$V'= 2\hbar N a \omega_{yz}  \fabsq{\psi (x)}/({1 + \omega_{yz}^2 t^2})$.
%
%
\begin{figure}[t]
\begin{center}
\includegraphics[angle=-90,width=1.\linewidth]{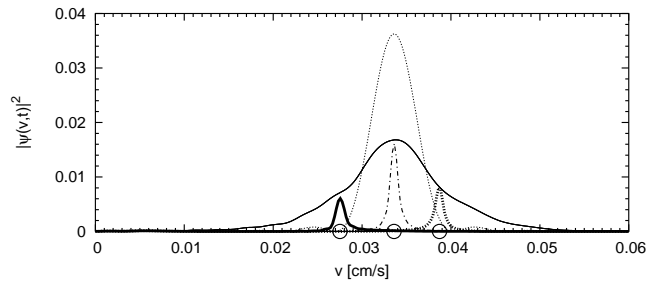}
\end{center}
\caption{\label{fig5} $|\psi(v, t)|^2$ for $v_0=0.0336\cms$,
$x_{TRAP} = -600 \mum$; $\widehat{V}_b = 300\hbar\si$,
$\omega_x = 5\si$, $\omega_{yz}=100\si, N = 5\times 10^4$;
$t=0$ (dotted line);
$t=0.8s$: solid line;
$t=8s$: $\widehat{V}_w = 140 \hbar\si$ (thick dotted line),
$\widehat{V}_w = 150 \hbar\si$ (dashed line),
$\widehat{V}_w = 160 \hbar\si$ (thick solid line);
the circles mark the resonance velocities
$v_R$.}
\end{figure}
%
Fig. \ref{fig5} shows the momentum distribution at $t=0$ (ground state). At $t=0.8 \, s$ 
the non-linearity has practically
vanished, so the momentum distribution stays stable 
until the velocity selection. 
The filtered distributions 
at $t=8 \, s$ for several 
resonance velocities
obtained with different well-depths are also shown.

We have in summary proposed an improvement 
of Fabry-Perot cavities to select the velocity of 
ultra-cold atoms using a well between the partially
reflecting mirrors,
and have provided simple formulae to explain and describe their
behaviour and the minimal velocity peak (zero) and width 
(non-zero) 
that can be achieved. 

\section*{Acknowledgments}

This work is supported by 
``Ministerio de Ciencia y Tecnolog\'\i a-FEDER''
(BFM2003-01003), and
UPV-EHU (Grant 15968/2004). AR acknowledges support of the German
Academic Exchange Service (DAAD)
and Ministerio de Educaci\'on y Ciencia.

\end{document}